\documentclass[11pt, a4paper]{article}

\usepackage{amsmath, amsthm, amssymb, amsfonts}
\usepackage{lmodern} 
\usepackage[utf8]{inputenc}
\usepackage{mathtools}
\usepackage{hyperref}
\usepackage{graphicx}
\usepackage{dsfont}
\usepackage{fullpage}
\usepackage{color}
\usepackage{soul} 
\usepackage[title]{appendix}
\usepackage{tikz}
\usetikzlibrary{patterns}
\usetikzlibrary{shapes.multipart}
\usetikzlibrary{arrows}
\usepackage{braket}

\definecolor{labelkey}{cmyk}{.4,.2,0,0}

\newcommand{\be}{\begin{equation}}
\newcommand{\ee}{\end{equation}}
\newcommand{\bea}{\begin{eqnarray}}
\newcommand{\eea}{\end{eqnarray}}

\newcommand{\R}{\ensuremath{\mathbb{R}}}

\newcommand{\Z}{\ensuremath{\mathbb{Z}}}

\renewcommand{\rho}{\varrho}

\newcommand{\eps}{\varepsilon}
\renewcommand{\leq}{\leqslant}
\renewcommand{\geq}{\geqslant}

\newcommand{\hd}{\mathsf{d}}
\newcommand{\he}{\mathsf{e}}
\newcommand{\hdo}{\mathsf{do}}
\newcommand{\heo}{\mathsf{eo}}

\usepackage{titlesec}
\titleformat{\section}{\large\bf}{\thesection}{1em}{}
\titleformat{\subsection}[runin]{\bf}{\thesubsection}{1em}{}[.]
\titleformat{\subsubsection}[runin]{\it}{\thesubsubsection}{1em}{}[.]

\theoremstyle{definition}

\theoremstyle{definition}

\theoremstyle{definition}

\theoremstyle{definition}
\newtheorem{remark}{Remark}[section]

\usepackage{authblk}
\author[1]{Guillaume Barraquand and Pierre Le Doussal}
\affil[1]{\normalsize Laboratoire de Physique de l'\'Ecole Normale Sup\'erieure, ENS, Universit\'e PSL, CNRS, Sorbonne Universit\'e, Universit\'e Paris-Cité, 75005 Paris, France}

\title{\bf \large Stationary measures of the KPZ equation on an interval from Enaud-Derrida's matrix product ansatz representation} 
\date{}

\begin{document}

\maketitle

\begin{abstract} 
The stationary measures of the Kardar-Parisi-Zhang equation on an interval have been computed recently. We present a rather direct derivation of this result by taking the weak asymmetry limit of the matrix product ansatz for the asymmetric simple exclusion process.  We rely on 
the matrix product ansatz representation of Enaud and Derrida,  which allows to express the steady-state in terms of re-weighted simple random walks. In the continuum limit, its measure becomes a path integral (or re-weighted Brownian motion) of the form encountered in Liouville quantum mechanics,
 recovering the recent formula. 

\end{abstract}


\section{Introduction }

\subsection{KPZ equation on an interval and its stationary measures} 

The Kardar-Parisi-Zhang (KPZ) equation \cite{KPZ} for the height field $h(x,t)$ on the interval $[0,L]$ is defined, for $0<x<L$, by the stochastic partial differential equation 
\be \label{eq:KPZ}
\partial_t h(x,t) = \partial_x^2 h + (\partial_x h)^2 + \sqrt{2} \xi(x,t)
\ee
where $\xi(x,t)$ is a standard space-time white noise. We impose  boundary conditions 
\begin{equation} 
\partial_x h\Big\vert_{x=0} = u, \quad \quad \partial_x h\Big\vert_{x=L} = -v,
\label{eq:boundary}
\end{equation} 
depending on two boundary parameters $u,v\in \R$. 
The solution is defined through the Cole-Hopf mapping $h(x,t)=\log Z(x,t)$,
where $Z(x,t)$ is the partition sum of a continuum directed polymer, which satisfies the stochastic
heat equation with Robin boundary conditions
$\partial_x Z(x,t)\vert_{x=0}=A Z(0,t)$ and $\partial_x Z(x,t)\vert_{x=L}=- B Z(L,t)$ with $u=A+1/2$ and $v=B+1/2$.
Although $Z(x,t)$ is not differentiable, the standard way to understand these boundary conditions is to impose 
these conditions on the heat kernel \cite{corwin2016open}, or through  a path integral as in 
\cite{borodin2016directed,deNardisPLDTT}. 

We will denote the stationary state by $\left(  H(x) \right)_{x\in [0,L]}$. It is stationary in the sense that if for a certain time $t_0$, the increment process $h(x,t_0)-h(0,t_0)$ has the same distribution as $H(x)$, this remains true for all times $t>t_0$. It is natural to expect that for fixed parameters $u,v$, the stationary measure is unique, and that for any initial condition $h(x,0)$, the increment process $h(x,t)-h(0,t)$ converges to $H(x)$ as $t$ goes to infinity.

The stationary measures of the KPZ equation 
on an interval were understood only recently. They have been first characterized in \cite{corwin2021stationary} and explicitly described in \cite{bryc2022markov, barraquand2022steady}, based on \cite{corwin2021stationary}. 
In particular, it was found in \cite{barraquand2022steady} that the distribution of $H$ admits a simple expression in terms of reweighted Brownian paths.  
More precisely, $H$ is the sum of two independent random fields
\be \label{eq:H} 
H(x) = \tfrac{1}{\sqrt{2}} W(x) + X(x)
\ee  
where $W(x)$ is a one-sided standard Brownian motion (i.e. with diffusion coefficient $1$, $W(0)=0$ and $W(L)$ free) 
and the probability distribution of the process $X(x)$ is given by the path integral measure 
\be  \label{eq:path} 
\frac{\mathcal D X}{{\cal Z}_{u,v}} e^{ -  \int_0^L dx  \left(\frac{dX(x)}{dx}\right)^2  }  e^{-2 v X(L)}\left( \int_0^L \mathrm dx\; e^{-2 X(x)} \right)^{-(u+v)}
\ee 
with $X(0)=0$ and $X(L)$ free. ${\cal Z}_{u,v}$ is a normalization such that ${\cal Z}_{0,0}=1$. In mathematical terms, $X$ is a continuous stochastic process on $[0,L]$ whose measure is absolutely continuous with respect to that of a Brownian motion with diffusion coefficient $1/2$  starting at $X(0)=0$, with Radon-Nikodym derivative 
\begin{equation} 
\label{eq:RNDforX}
\frac{1}{\mathcal Z_{u,v}} \left( \int_0^L \mathrm dx e^{-2X(x)}\right)^{-u}  \left(\int_0^L \mathrm dx e^{2X(L)-2X(x)}\right)^{-v}.
\end{equation} 
This equivalent form makes apparent that the process is invariant 
after reversing space and exchanging $u,v$.

 The distribution of $H$ given above was obtained via a Laplace inversion of an explicit formula for the multipoint distribution of the field $H$, obtained
in \cite{corwin2021stationary}. Although \cite{corwin2021stationary} restricted to $u+v>0$, and the Laplace transform  formula therein makes sense only in that case, the measure \eqref{eq:path} makes sense for any value of $u,v\in \R$. This led us to  conjecture in \cite{barraquand2022steady} that \eqref{eq:H} is the stationary measure for all values of $u$ and $v$. When $u+v>0$, another equivalent description of the law of $H$ was obtained slightly earlier in \cite{bryc2022markov}, also based on \cite{corwin2021stationary}, in terms of a Doob-like transform of a process with an explicit transition kernel. Although the Laplace inversions in  \cite{bryc2022markov} and \cite{barraquand2022steady} were performed independently, they are quite similar for the following reason: 
When $u+v>0$, the process $X$ can be written as $X(x)=U(x)-U(0)$, where the distribution of $U$ is given by the path integral measure \cite{barraquand2022steady} 
\begin{equation}
\label{eq:PU}
    \frac{\mathcal D U}{\widetilde{\mathcal Z}_{u,v}} \exp\left(-2u U(0)-2vU(L)  - \int_{0}^{L} dx \left[ \left(\frac{dU(x)}{dx}\right)^2 +  e^{-2 U(x)}\right] \right),
\end{equation}
where the endpoints $U(0)$ and $U(L)$ are now both free. 
Hence, $U(x)$ obeys Liouville quantum mechanics  on $x \in [0,L]$ with some specific boundary weights for $U(0)$ and $U(L)$. 
The path integral \eqref{eq:PU} was obtained as an intermediate step in \cite{barraquand2022steady} and one deduces \eqref{eq:path} from \eqref{eq:PU} by simply integrating over $U(0)$, the  ``zero mode'' of the Liouville action -- a standard procedure in the Liouville theory \cite{goulian1991correlation,TexierComtetSUSY}. On the other hand, the Green's function associated with the Liouville Hamiltonian $\frac{-1}{4} \frac{d^2}{dU^2} + e^{-2U}$ is what is called the Yakubovich heat kernel in the reference \cite{bryc2022markov}, where the process $U$ (denoted $Y$ in \cite{bryc2022markov}) is defined through its Markovian transition densities. Using the Feynman-Kac theorem (see \cite[Eq. (3.4)]{bryc2022markov} or \cite[Eq. (16)]{barraquand2022steady}), this shows that the descriptions appearing in \cite{bryc2022markov} and  \cite{barraquand2022steady} are equivalent, at least for $u,v>0$. The equivalence was proved to hold further for any $u,v$ such that $u+v>0$ in \cite{bryc2021markov}. 
 Let us also mention that defining the process $U(x)$ through a Markov process having an explicit transition kernel rather than through a path integral also has certain advantages, for instance to compute observables of the process \cite[Supplementary Material]{barraquand2022steady} or to justify rigorously certain limits \cite{bryc2021markov}. However, in this paper, we will focus on the path integral (or Brownian reweighting) point of view.

\subsection{Aim of this paper} 
There does not exist any general method to compute stationary measures of non-linear non-equilibrium  stochastic PDEs such as the KPZ equation. We are not even aware of any method to directly check that the process defined by   \eqref{eq:H} and \eqref{eq:path} is stationary on the interval with the boundary conditions \eqref{eq:boundary}. However, it is well-known that the KPZ equation can be approximated by various discrete integrable models, whose stationary measure is sometimes explicit. The stationary measure of the KPZ equation can hence be obtained as an appropriate scaling limit of the steady-state of the discrete model. For the KPZ equation on an interval, the only good candidate seems to be the asymmetric simple exclusion process (ASEP) whose steady state  was obtained in terms of the matrix product ansatz (MPA) \cite{derrida1993exact}. The convergence of ASEP's height function to the KPZ equation in the weak asymmetry limit was first proved in \cite{bertini1997stochastic} for the case of the KPZ equation on $\R$, and in \cite{corwin2016open} for the case of the KPZ equation on an interval, which we are interested in here. 

Nevertheless, the description of the stationary measures for the KPZ equation  on an interval was obtained as a combination of a long series of highly nontrivial works. It is based on a specific representation of the MPA from \cite{uchiyama2004asymmetric} involving Askey-Wilson orthogonal polynomials. In an apparently disconnected area of mathematics, these Askey-Wilson orthogonal polynomials were used to construct stochastic processes called quadratic harnesses in \cite{bryc2010askey}. The connection with the MPA was made in \cite{bryc2017asymmetric}, where it was shown that  the Laplace transform of ASEP's stationary measure can be written using some of those quadratic harnesses. The weak asymmetric scaling limit was then considered in \cite{corwin2021stationary}, which established the first formula characterizing stationary measures for the KPZ equation on an interval. The multipoint Laplace transform of the stationary measure was expressed in \cite{corwin2021stationary} as multiple integrals of products of ratios of Gamma functions. Finally, \cite{barraquand2022steady} recognized in the formulas from \cite{corwin2021stationary} expressions that typically arise in Liouville quantum mechanics and performed a Laplace inversion to obtain  \eqref{eq:H} and \eqref{eq:path}. The alternative Laplace transform inversion from \cite{bryc2022markov} uses similar ideas, though rather inspired by results in stochastic analysis  than Liouville quantum mechanics.  

The aim of this paper is to show that there exists a more direct derivation of the stationary measure \eqref{eq:H} via the path integral \eqref{eq:PU}, as was suggested to us by Bernard Derrida. This alternative route relies on two main inputs. The first one is a representation of the MPA, due to Enaud and Derrida \cite{enaud2004large}, different from the representation from \cite{uchiyama2004asymmetric}. The second input is that  when the MPA representation is made of reasonably simple bidiagonal matrices, it is possible to describe the stationary height function as some reweighted random walk. This elegant trick was first used in \cite{derrida2004asymmetric} in the totally asymmetric case (TASEP), and in \cite{enaud2004large, derrida2005fluctuations} in the partially asymmetric case (ASEP). Connections between the MPA and Motzkin paths are also discussed in \cite{brak2006combinatorial, blythe2009continued}, though these works involve different representations of the MPA, hence the weighted lattice paths there are different from the present paper. Our main result is that in the weakly asymmetric scaling limit, 
the reweighted random walks defined below in Section \ref{sec:ASEP} converge to the reweighted Brownian motion defined by \eqref{eq:PU}, ultimately leading to the representation \eqref{eq:H}. This new derivation is close in spirit to the arguments used in \cite{enaud2004large}, although this reference considered a different scaling limit of ASEP which does not converge to the KPZ equation and studied large deviations. 

\subsection{Some open questions} 
As  mentioned above, we conjectured in \cite{barraquand2022steady} that \eqref{eq:H} and \eqref{eq:path} hold for any value of $u,v\in \R$. The conjecture was based on the fact that we expect that for any finite $L$, the multipoint density, and other observables of the stationary process, should be analytic in parameters $u$ and $v$. Since the Radon-Nikodym derivative in \eqref{eq:RNDforX} is analytic, we expect that the result can be extended from $u+v>0$ to the whole $u,v$ plane.   The measure of the field $U$ in \eqref{eq:PU} however, makes sense only when $u+v>0$, hence the new derivation that we present in this paper works only under this condition (the MPA representation that we start from is valid only under a similar condition). We believe that it would be interesting to investigate other representations of the MPA, to find a derivation of KPZ stationary measures that would work for $u+v<0$. By considering the $L\to\infty$ limit of the KPZ stationary measures on $[0,L]$ and using stochastic analysis results of \cite{hariya2004limiting}, we also conjectured in \cite{barraquand2022steady} a description of the stationary measures of the KPZ equation on the half-line $\R_+$. It turns out that the latter have been proved rigorously in \cite{barraquand2022stationary}, using another discretization of the KPZ equation (the log-gamma polymer), and this without assuming $u+v>0$. This provides more weight towards the conjecture that \eqref{eq:H} and \eqref{eq:path} hold for any value of $u,v\in \R$. 

The derivation of KPZ stationary measures in the present paper is not completely mathematically rigorous. The first step is to express ASEP's height function, under the stationary measure, as the sum of two random walks $\vec n$ and $\vec m$, see Section \ref{sec:ASEP}, and this poses no issue. Then, we consider the scaling limit. Writing the asymmetry in ASEP as $q=e^{-\eps}$, letting $\eps$ to zero, and scaling the random walks appropriately, we argue in Section \ref{sec:KPZ} that the random walks converge to a couple of (reweighted) Brownian motions $(U,V)$ with $U$ distributed as in \eqref{eq:PU} and $V$ distributed as $\frac{1}{\sqrt{2}}W$ in \eqref{eq:H}. However, we have not fully justified that the sequence of rescaled random walks $(U_{\eps}, V_{\eps})$ (defined in \eqref{eq:defUeps} below) built from $(\vec n, \vec m)$ converges to $(U,V)$. We only identified the limit, assuming the tightness of the sequence of processes  $(U_{\eps}, V_{\eps})$. Proving tightness remains an open mathematical problem.

\subsection{Outline of the paper} 
In Section \ref{sec:ASEP}, we recall the definition of the open ASEP and the description of its steady-state using the MPA. We also give an interpretation of the steady state in terms of the sum of two reweighted random walks, following \cite{derrida2004asymmetric,enaud2004large}. In Section \ref{sec:KPZ}, we first recall the scaling limit of ASEP height function to a solution of the KPZ equation \cite{corwin2016open}. We then explain how the reweighted random walks of the MPA
converge to reweighted Brownian motions leading to our main result for the stationary measure of the KPZ equation on the interval. 

\section{Matrix product ansatz for ASEP and sum over paths}
\label{sec:ASEP}
\subsection{Definition of open ASEP} 

Let us first recall the definition of the asymmetric exclusion process (ASEP) with open boundaries. We will use here the notations of \cite{enaud2004large} 
with the only change that $L$ there is denoted $\ell$ here. The system is a one dimensional lattice gas on $\ell$ sites. At any given time, 
each site $1 \leq i \leq \ell$ is either empty or occupied by at most one particle. A configuration of particles is described by a collection of occupation numbers $( \tau_i )_{1\leq i\leq \ell}$ where $\tau_i=1$ when site $i$ is occupied and 
$\tau_i=0$ when site $i$ is empty. The model depends on bulk parameters $p$ and $q$ and boundary parameters $\alpha, \beta, \gamma, \delta$. 
 The dynamics is a continuous time Markov process $\tau(t)$ on the space state $( \tau_i )_{1\leq i\leq \ell} \in \{0,1\}^\ell$ defined as follows: 
 
At any given time $t>0$ and for any $i$, $1 \leq i \leq \ell-1$, each particle jumps from $i$ to $i+1$ with exponential rate $p \tau_{i}(1-\tau_{i+1})$
and from $i+1$ to $i$ with exponential rate $q \tau_{i+1}(1-\tau_{i})$. In addition a particle at site $1$ is created 
with exponential rate $\alpha (1- \tau_1)$ and annihilated with exponential rate $\gamma \tau_1$, and a particle at site $\ell$ is created 
with exponential rate $\delta (1- \tau_\ell)$ and annihilated with exponential rate $\beta \tau_\ell$. All these events are independent.

We introduce density  parameters 
$\rho_a, \rho_b$ which are related to the boundary rates $\alpha,\beta,\gamma,\delta$ by the mean current conservation
conditions $j_a=(p-q) \rho_a (1-\rho_a)=\alpha(1-\rho_a)-\gamma \rho_a$ and 
$j_b=(p-q) \rho_b (1-\rho_b)=\beta \rho_b -\delta (1-\rho_b)$ \cite{derrida1993exact}. 

We will restrict to parameters $\alpha, \beta, \gamma, \delta$ solving Liggett's condition \cite{liggett1975ergodic}
\begin{equation}
    \frac{\alpha}{p}+\frac{\gamma}{q}=1, \;\;\; \frac{\beta}{p}+\frac{\delta}{q}=1,
    \label{eq:Liggett}
\end{equation}
so that 
the density parameters are related to the jump rates by 
\begin{equation} \label{eq:rho} 
\rho_a=\frac{\alpha}{p},\;\; \rho_b=\frac{\delta}{q}.
\end{equation}
Thus, assuming \eqref{eq:Liggett} does not restrict the range of density parameters $\rho_a, \rho_b \in [0,1]$ and allows to access the full phase diagram.  
 
The interpretation of the conditions \eqref{eq:Liggett} (explained in  \cite{liggett1975ergodic}) is that at the left boundary, the injection rate $\alpha$ can be written as $\alpha=p\rho_a$, while the rate $\gamma$ can be written as $\gamma= q(1-\rho_a)$ as if the behaviour of the reservoir could be replaced by a fictitious site $0$, occupied with probability $\rho_a$, communicating with the rest of the system as in the bulk. Similarly, at the right boundary, the injection rate is $\delta=\rho_b q$, while the ejection rate is $\beta=p(1-\rho_b)$, with a similar interpretation involving a fictitious site $\ell+1$.

\subsection{Matrix product ansatz} 
The stationary probability measure $P(\tau)$, where $\tau=(\tau_i)_{1\leq i\leq \ell}$, is given by the matrix product ansatz \cite{derrida1993exact}
\be \label{eq:MPA}
P(\tau) = \frac{1}{Z_\ell(q)} \langle W | \prod_{i=1}^\ell ( D \tau_i + E (1- \tau_i) ) | V \rangle, \quad  \quad 
Z_\ell(q)= \langle W | ( D + E )^\ell | V \rangle.
\ee 

For stationarity to hold,  the matrices $D$ and $E$, as well as the ket $|V \rangle$ and bra $\langle W|$
must satisfy the following algebraic relations \cite{derrida1993exact}:
\begin{subequations}
    \begin{align} 
     pDE-qED&=D+E,\label{eq:MPA1}\\ 
     (\beta D-\delta E) \ket{V} &=\ket{V}\label{eq:MPA2},\\
     \bra{W} (\alpha E-\gamma D)&=\bra{W}\label{eq:MPA3}.
    \end{align}
    \label{eq:MPArelations}
\end{subequations}
The normalization constant is then related to the current in the stationary state $j=Z_{\ell-1}(q)/Z_\ell(q)$.

There are various representations of these relations \cite{derrida1993exact,  sandow1994partially, essler1996representations, sasamoto1999one, blythe2000exact, enaud2004large, uchiyama2004asymmetric}  and here we use the one given in \cite{enaud2004large} -- see Appendix \ref{sec:appendixMPA} for details about this representation,
which we now recall. First of all, one sets $p=1$, without loss of generality. Furthermore, the representation found in \cite{enaud2004large} 
is valid for any values of $\alpha, \beta ,\gamma, \delta$, and is parameterized by the two density parameters $\rho_a, \rho_b$, as well as two extra parameters $d$ and $e$ defined in \cite{enaud2004large} (see Appendix \ref{sec:appendixMPA}). Imposing the condition \eqref{eq:Liggett} corresponds to letting $e=d=q$, which simplifies slightly the representation. 
We will choose $D$ and $E$ as the infinite matrices
\be \label{eq:DEmat} 
D = \begin{pmatrix} 
[1]_q & [1]_q & 0 & 0 & 0 &\cdots \\ 
0 & [2]_q & [2]_q & 0 & 0 &\cdots \\
0 & 0 & [3]_q & [3]_q & 0 &\cdots \\
\vdots &  \vdots &  0  &\ddots     & \ddots &  \ddots
\end{pmatrix}, \quad  \quad 
E = \begin{pmatrix} 
[1]_q & 0 & 0 & 0 &  \cdots \\
[2]_q & [2]_q & 0 & 0 & \cdots \\
0 & [3]_q & [3]_q & 0 &  \\
0  & 0  & \ddots  & \ddots   & \ddots \\

\end{pmatrix} 
\ee 
where we use the notation 
\be 
[n]_q=\frac{1-q^n}{1-q}.
\ee 
Let us denote by $\{|n \rangle \}_{n \geq 1}$ the vectors of the associated basis. One can alternatively write
the matrices $D$ and $E$ as 
\begin{equation} 
D= \sum_{n=1}^{+\infty}  [n]_q \ket{n} (  \bra{n}+ \bra{n+1}) , \quad \quad
E= \sum_{n=1}^{+\infty}  ([n]_q\ket{n}+[n+1]_q\ket{n+1}) )\bra{n}.
\end{equation} 
In this basis the vectors $|V \rangle$ and $\langle W|$ are given by
\be \label{eq:defWV}
\langle W | = \sum_{n \geq 1} \left(\frac{1-\rho_a}{\rho_a}\right)^n \bra n , \quad  
\quad 
|V \rangle = \sum_{n \geq 1} \left(\frac{\rho_b}{1-\rho_b}\right)^n [n]_q \ket n.
\ee

\begin{remark} \label{remark1} It is very useful for the following to note that one can rewrite 
$D,E$ as $D=\Lambda \widetilde D $, $E=\Lambda \widetilde E$ where $\Lambda$ is the diagonal matrix 
\begin{equation}
    \Lambda=\sum_{n=1}^{+\infty} [n_q] \ket n\bra n, \quad \widetilde D=\sum_{n=1}^{+\infty}  \ket{n} (  \bra{n}+ \bra{n+1}), \quad \widetilde E=\sum_{n=1}^{+\infty}  (\ket{n}+\ket{n+1}) )\bra{n},
\end{equation} 
that is all the entries $[n]_q$ in \eqref{eq:DEmat} are replaced by $1$. The matrices $\widetilde D,\widetilde E$ are
the same matrices as in the case of TASEP \cite{derrida2004asymmetric} (see also \cite{derrida1993exact}),
which corresponds to $q=0$. 
\end{remark} 

\begin{remark} 
In order for the distribution \eqref{eq:MPA} to be well-defined, one needs that the normalization $Z_{\ell}(q)<\infty$. 
For the representation that we have chosen,  when $q\in [0,1)$, it is easy to see that $\langle W \vert V \rangle <\infty$ is equivalent to $\rho_b<\rho_a$ (which becomes $u+v>0$ in the scaling limit).  This condition is sufficient since  $\langle W \vert V \rangle <\infty$ implies that for any fixed $\ell$, the normalization constant $Z_{\ell}(q) = \bra{W} (D+E)^{\ell}\ket{V} $ is finite as well. The distribution is also well-defined for $q>1$ under some extra condition, see \cite{enaud2004large}.
\label{rem:finiteZ}
\end{remark}  

\subsection{Weighted random walks} From now on we assume that $q\in [0,1)$ and $\rho_b<\rho_a$. 
Following \cite{derrida2004asymmetric} and \cite{enaud2004large} we introduce discrete walks on the strictly positive integers which, at each step, either increase by one unit, decrease by one unit, or stay constant, with the constraint that they remain strictly positive. They are described by a sequence of $\ell+1$ integers $ \vec n=(n_i)_{0\leq i\leq \ell}$ such that 
$n_i >0$ and $|n_i-n_{i+1}| \leq 1$ for all $0\leq i\leq \ell-1$. To each walk $\vec n$ one associates the weight $\Omega(\vec n)$ defined as
\be \label{Omega} 
\Omega(\vec n) = \left(\frac{1-\rho_a}{\rho_a}\right)^{n_0} \left(\frac{\rho_b}{1-\rho_b}\right)^{n_{\ell}} \prod_{i=1}^{\ell} v(n_{i-1}, n_i)\prod_{i=0}^{\ell} [n_i]_q, 
\ee 
where 
\be 
v(n,n')= \begin{cases} 2 &\mbox{ if }n=n',\\ 1 &\mbox{ if } \vert n-n'\vert=1 \\ 
0 &\mbox{ else}.
\end{cases} 
\ee 
One can first check that the normalization can be written as 
\be
Z_\ell(q)=\bra{W} (D+E)^\ell \ket{V} = \sum_{\vec n} \langle W |n_0 \rangle \langle n_\ell |V \rangle \prod_{i=1}^\ell \langle n_{i-1} | D + E | n_{i} \rangle 
=
\sum_{\vec n} \Omega(\vec n) ,
\ee
where we used Remark \ref{remark1} to factor out the product of $[n_i]_q$.

Next, we shall think of $\vec n$ as a random walk and define $\nu$ to be the measure on walks $\vec n$ such that 
\be 
\nu(\vec n) = \frac{\Omega(\vec n)}{\sum_{\vec n} \Omega(\vec n)}.
\ee

One can generate steady state configurations of the occupation
numbers $\tau_i$ from the random walk $\vec n$. This is explained in \cite{derrida2004asymmetric}
for the totally asymmetric case, that is when $q=0$. The main property which is needed is that 
\be 
\langle n | D | n' \rangle = \frac{1 + n'-n}{2} \langle n | D + E | n' \rangle.
\ee 
One can check that this property still holds for $q>0$ since the factor $[n]_q$ is in factor on each side of the equation. 
It allows to express the multi-point correlation  (which determine the distribution of $\tau$) as
\be 
\langle \tau_{i_1} \dots \tau_{i_k} \rangle = \sum_{\vec n} \nu(\vec n) \prod_{j=1}^k \frac{1 + n_{i_j} - n_{i_j-1}}{2},
\ee 
for any $i_1<\dots<i_k$,  where $\langle \cdot \rangle$ denotes the expectation with respect to the probability measure $P$. 
This implies \cite{derrida2004asymmetric} that the joint distribution $\nu(\vec n,\tau)$ of $\vec n$ and $\tau=(\tau_i)_{1\leq i\leq \ell}$ has the form
$\nu(\vec n,\tau)=\nu(\vec n)P(\tau|\vec n)$ where the conditional distribution of $\tau$ given $\vec n$ is 
\begin{equation} \label{conditional}
P\left(\tau\vert\vec n\right) = \prod_{i=1}^\ell \frac{1 + (n_i-n_{i-1}) (2 \tau_i-1) }{2}.
\end{equation}
Hence, as pointed out in \cite{derrida2004asymmetric}, there is a simple way to generate steady state configurations of the occupation
numbers $\tau_i$. First one generates a random walk $\vec n$ with $\ell$ steps according to the measure $\nu(\vec n)$.
Then a steady state configuration $\tau=(\tau_i)_{1 \leq i\leq \ell}$ is obtained by taking $\tau_i= 1$ whenever $n_i - n_{i-1} = 1$,
taking 
$\tau_i=0$ whenever $n_i - n_{i-1} = -1$, and choosing $\tau_i=0$ or $\tau_i=1$ with equal probability for
each $i$ such that $n_i - n_{i-1} = 0$.

Thus, under the stationary measure $P(\tau)$, we may write the process $\tau$ as the sum of two processes
\be 
2\tau_i -1 = n_i-n_{i-1} + \sigma_i 
\label{eq:discretesum}
\ee
where $\vec n$ is distributed according to the measure $\nu$, and conditionally on $\vec n$, the $\sigma_i$ are independent random variables  equal to $0$ when $n_{i-1} \neq n_i$, and equal to $\pm 1$ with equal probability $1/2$ otherwise.  

Letting $m_i=\sum_{j=1}^i\sigma_i$, one notices that $(n_i,m_i)$ is just a two-dimensional simple random walk on $\mathbb Z^2$ reweighted by a functional of $\vec n$.  Indeed, one can rewrite \eqref{eq:discretesum} as follows.  Under the stationary measure $P(\tau)$, 
\begin{equation}  \left(\sum_{j=1}^i (2\tau_i-1)\right)_{1\leq i\leq \ell} = \left(n_i-n_0+ m_i \right)_{1\leq i\leq \ell},
\label{eq:heightincrements}
\end{equation}
where $(n_i, m_i)_{0\leq i\leq \ell}$ is a two dimensional random walk on $ \Z^2$, starting from $(n_0,0)$, distributed as 
\begin{equation} P(\vec n, \vec m) = \frac{\mathds{1}_{n_0>0}}{4^{-\ell} Z_{\ell}(q)}\left(\frac{1-\rho_a}{\rho_a}\right)^{n_0} \left(\frac{\rho_b}{1-\rho_b}\right)^{n_{\ell}} \prod_{i=0}^{\ell} [n_i]_q P^{SSRW}_{n_0,0}(\vec n, \vec m), 
\label{eq:measurenm}
\end{equation}
where $P^{SSRW}_{n_0,0}$ denotes the probability measure of the symmetric simple random walk (SSRW) on $\mathbb Z^2$ starting from $(n_0,0)$, that is the random walk performing steps $(1,0)$, $(0,1)$, $(-1,0)$ and $(0,-1)$ with equal probability $1/4$. Note that in \eqref{Omega}, the random walk $n_i$ was constrained to the strictly positive integers. However, the weight $\Omega(\vec n)$ vanishes if $n_i=0$ for some $i$, so that one can relax the positivity assumption. 

\subsection{Summing over the zero mode}
As in \cite{barraquand2022steady} in the context of open KPZ stationary measures, it is possible to sum over $n_0$ in \eqref{eq:measurenm}. Let us introduce variables $x_i=n_i-n_0$ for $0\leq i\leq \ell$. Observe that 
\begin{equation}
	 \prod_{i=0}^{\ell} [n_i]_q = (1-q)^{-\ell-1} \sum_{k=0}^{\ell} (-1)^k q^{n_0k} e_k\left( \left\lbrace q^{x_i}\right\rbrace_{0\leq i\leq \ell}\right), 
\end{equation}
where the $e_k$ denote the elementary symmetric polynomials $e_k(z_1, \dots, z_n)= \sum_{i_1<\dots<i_k}z_{i_1}\cdots z_{i_k}$. Hence, using the shorthand notation  $R= \frac{1-\rho_a}{\rho_a}\frac{\rho_b}{1-\rho_b}$, 
\begin{equation}
	\sum_{n_0=1}^{+\infty} P(\vec n, \vec m) = \frac{(1-q)^{-\ell-1}}{4^{-\ell} Z_{\ell}(q)} \left(\frac{\rho_b}{1-\rho_b}\right)^{x_{\ell}}\sum_{k=0}^{\ell} \frac{(-1)^k}{1-q^k R} e_k\left( \left\lbrace q^{x_i}\right\rbrace_{0\leq i\leq \ell}\right) P^{SSRW}_{0,0}(\vec x, \vec m).
	\label{eq:aftersummation}
\end{equation} 
This probability measure on the random walk $(x_i, m_i)_{0\leq i\leq \ell}$ now makes sense for any values of $\rho_a, \rho_b\in (0,1)$, without imposing the condition $\rho_b<\rho_a$. However, unlike what happens for the open KPZ equation stationary measures \cite{barraquand2022steady}, the form \eqref{eq:aftersummation} is  more complicated than \eqref{eq:measurenm}. Furthermore, we do not expect \eqref{eq:aftersummation} to describe ASEP stationary measure when $\rho_b>\rho_a$.

\begin{remark}When $R\to 1$, that is when $\rho_a\to\rho_b$, only the term $k=0$ dominates in \eqref{eq:aftersummation} and we see that the random walk $(x_i, m_i)_{0\leq i\leq \ell}$ becomes a simple random walk on $\Z^2$ making steps $(0,1)$ and $(0,-1)$ with probabilities $\rho_b(1-\rho_b)$, and steps $(1,0)$ and $(-1,0)$ with probabilities  $\rho_b^2$ and $(1-\rho_b)^2$. The random walk $(x_i+m_i)_{0\leq i\leq \ell}$ defined in \eqref{eq:heightincrements} is then a simple random walk with probabilities $\rho_b$ and $1-\rho_b$. Hence, we recover that in the limit $\rho_a\to \rho_b$, the stationary measure becomes product Bernoulli($\rho_b$).
\end{remark}

\section{Scaling limit from ASEP to the KPZ equation}
\label{sec:KPZ}

\subsection{Height function in ASEP}
We may slightly extend the definition of ASEP by considering it as a Markov process on $\lbrace 0,1\rbrace^{\ell}\times \Z$, described by occupation numbers $\tau\in \lbrace 0,1\rbrace^{\ell}$, as well as the net number $N\in \Z$ of particles that have entered the system, from the left reservoir to site $1$, minus the number of particles that have exited the system from site $1$ to the left reservoir since time $t=0$.  On this extended state space, we define a height function $h_t(i)$, following \cite{corwin2016open}, as 
\begin{equation}
    h_t(i) = -2N+ \sum_{j=1}^i (2\tau_j-1), \;\;\; h_t(0)=-2N. 
\end{equation}
By this definition, \eqref{eq:heightincrements} implies that under the stationary measure, the height function increments can be written as 
\begin{equation}
    h_t(i)-h_t(0) = n_i-n_0+ m_i,
    \label{eq:stationaryheightdiscrete}
\end{equation}
where $\vec n$ and $\vec m$ are distributed according to \eqref{eq:measurenm}.

\subsection{Convergence of ASEP to the KPZ equation}

Let us now describe the convergence of ASEP to the KPZ equation proved in \cite{corwin2016open}. We consider the scalings, for $\eps>0$, 
\begin{equation} 
q=e^{-\eps}, \quad  \quad i= 4 x\eps^{-2}, \quad \quad \ell= 4  L \eps^{-2}. 
\label{eq:scalings}
\end{equation}
We  also scale the boundary rates $\alpha, \beta, \gamma, \delta$ in such a way that 
\be 
\rho_a = \frac{1}{2} + \frac{u}{4}  \eps, \quad  \quad \rho_b = \frac{1}{2} - \frac{v}{4}  \eps.
\ee 
Recall that by Remark \ref{rem:finiteZ} we must assume that $\rho_a>\rho_b$, so that $u+v>0$. 
Then, we define $Z_t(j) = e^{\eps h_t(j)/2+c t}$, where $c=1+q-2\sqrt{q}$. Theorem 2.18 in \cite{corwin2016open} states that $Z_{16 \eps^{-4} T}(4\eps^{-2} x)$ weakly  converges as a space time continuous stochastic process, as $\eps $ goes to zero, to the process $Z(x,T)$, solution of the stochastic heat equation (SHE)
\begin{equation}
    \partial_T Z = \partial_{xx} Z + \sqrt{2} Z \xi(x,T).
\end{equation}
Note that in \cite{corwin2021stationary}, all rates are multiplied by $1/2$, compared to the scalings discussed just above, but this is consistent with the fact that our definition of the KPZ equation and the SHE also corresponds to scaling the time by $2$ compared to the convention in \cite{corwin2016open}. Note also that the parameter $\epsilon$ in \cite{corwin2021stationary} is not the same as our $\eps$. One needs to match the asymmetry $q/p=e^{-2\sqrt{\epsilon}}$ from \cite{corwin2016open} with the asymmetry $q=e^{-\eps}$ in the present paper, so that denoting by $\epsilon$ the parameter epsilon in \cite{corwin2016open} we must take $\eps= 2 \sqrt{\epsilon}$ to match notations.

\subsection{Scaling limit of the steady-state}
Let us define
\begin{equation}
H_{\eps}(x) = \frac{\eps}{2} \left(h_t(4x \eps^{-2} )-h_t(0)\right)  = \frac{\eps}{2} \left( n_{4x \eps^{-2}}-n_0 + m_{4x \eps^{-2}} \right), 
\end{equation}
where the random walks $\vec n, \vec m$ are distributed according to $P(\vec n, \vec m)$ defined in \eqref{eq:measurenm}. 
According to the convergence from \cite{corwin2016open}, if $\tau$ is distributed according to ASEP's steady-state, then $H_{\eps}(x)$ converges when $\eps$ goes to zero, as a process of the variable  $x$, to the stationary state of the KPZ equation denoted $H(x)$ above. The main idea is that in the continuum limit the random walk $\vec n$
will lead to the process $U(x)$, and  the random walk $\vec m$ will lead to the Brownian motion $\frac{W(x)}{\sqrt{2}}$. 
Finally, we will show that in the continuum limit the random walks $\vec n$ and $\vec m$ become independent processes so that  the increments of the height function $H_\eps(x)$ will lead to the process $H(x)= U(x) + \frac{W(x)}{\sqrt{2}}$. 

We define functions $U_{\eps}, V_{\eps}$ on $[0,L]\cap \frac{\eps^2}{4} \Z$ by
\be 
U_\eps(x) = \frac{\eps}{2} (n_{4x \eps^{-2}} + \eps^{-1} \log(\eps^2/4) ), \quad \quad  V_{\eps}(x) = \frac{\eps}{2} m_{4x \eps^{-2}},  
\label{eq:defUeps}
\ee
that we extend linearly to the whole interval $[0,L]$. Note that $U_{\eps}$ lives in the interval $(\log(\eps/2), +\infty)$ which becomes the whole line $\R$ in the $\eps\to 0$ limit.  
To determine the limit of $U_{\eps}, V_{\eps}$ as $\eps\to 0$, we need to  examine the behavior of \eqref{eq:measurenm} under this scaling.  
First, under the reference measure $P^{SSRW}_{n_0,0}$, the couple of random functions  $(U_{\eps}(x), V_{\eps}(x))$ becomes a two-dimensional Brownian motion.
 Each component of the SSRW $(n_i,m_i)$ evolves by $+1$ or $-1$ with probability $1/2$ at each step, and stays put with probability $1/2$. Since the diffusion coefficient is preserved by the scaling \eqref{eq:defUeps}, we find that under the measure $P^{SSRW}_{n_0,0}$, the couple  $(U_{\eps}(x), V_{\eps}(x))$ weakly converges to a two-dimensional Brownian motion $(U(x),V(x))$ with covariance matrix $\frac{1}{2}I_2$,  starting from $(U(0), 0)$.

Next we obtain  
\begin{align} 
    \prod_{i=0}^{\ell} [n_i]_q &= \exp\left( \sum_{i=0}^{\ell} \log\left(  \frac{1 - e^{- n_i \eps}}{1- e^{-\eps}} \right) \right)\\
    &= C(\eps) \exp\left(  \sum_{i=0}^{4 L\eps^{-2}} \log(1-e^{\log(\eps^2/4)- 2 U_{\eps}(\eps^2 i/4)}) \right)  \\
    &\simeq C(\eps) \exp\left(  -\sum_{i=0}^{4 L\eps^{-2}} \frac{\eps^2}{4} e^{- 2 U_{\eps}(\eps^2 i/4)} \right)  \\ 
    &\simeq C(\eps ) \exp\left( -\int_0^L e^{-2 U_{\eps}(x)}dx\right),
    \end{align}
where the constant  $C(\eps)=\eps^{-(\ell+1)}$, independent from $U_{\eps}$,
disappears when dividing by the normalization. 
 Under the same scalings, the boundary terms in \eqref{eq:measurenm} yield
\be 
\left(\frac{1-\rho_a}{\rho_a}\right)^{n_0} \left(\frac{\rho_b}{1-\rho_b}\right)^{n_{\ell}} \simeq C'(\eps) e^{- 2 u U_{\eps}(0) - 2 v U_{\eps}(L)}, 
\ee 
up to another unimportant multiplicative factor $C'(\eps)=(\eps^2/4)^{u+v}$, which also disappears when dividing by the normalization.

Thus, the couple of random function $(U_{\eps}(x), V_{\eps}(x))$ converge, under the scalings \eqref{eq:scalings}, to a two-dimensional process $(U(x),V(x))\in \R^2$, absolutely continuous with respect to the law of the two-dimensional Brownian motion with covariance matrix $\frac{1}{2}I_2$, where $V(0)=0$ and $U(0)$ is distributed according to the Lebesgue measure. The Radon-Nikodym derivative is equal to  
\begin{equation}
\frac{1}{\widetilde{\mathcal Z}_{u,v}} \exp\left(-2u U(0)-2vU(L)  - \int_{0}^{L} dx  e^{-2 U(x)}\right).
\label{eq:RNDcontinuum}
\end{equation}
We recover the formula \eqref{eq:PU} for the measure of the field $U(x)$ with boundary parameters $u,v$.
Furthermore, since the Radon-Nikodym derivative \eqref{eq:RNDcontinuum} does not involve $V$, the two components $U$ and $V$ remain independent despite the reweighting, so that 
\eqref{eq:H} holds (with $V=\frac{1}{\sqrt{2}}W$).

\subsection*{Acknowledgments}
We thank B. Derrida for an enlightening discussion about the methods of \cite{derrida2004asymmetric} and \cite{enaud2004large}. 
P. Le Doussal was supported by ANR grant ANR-17-CE30-0027-01 RaMaTraF. G. Barraquand was supported by ANR grant ANR-21-CE40-0019. We also acknowledge hospitality and support from Galileo Galilei Institute, during the participation of both authors to the scientific program on “Randomness, Integrability, and Universality” in the Spring 2022.

\newpage 
\appendix
\begin{center}
    \large\bf  Appendix
\end{center}
\renewcommand\thesection{\Alph{section}}

\section{Matrix Product Ansatz representation}
\label{sec:appendixMPA}
In this Section, we explain how to find a representation of the MPA relations \eqref{eq:MPA}, and in particular, how to obtain \eqref{eq:DEmat} and \eqref{eq:defWV}, following \cite{enaud2004large}. The representation found in \cite{enaud2004large} is more general than the one we used above in Section \ref{sec:ASEP}, as it does not assume that \eqref{eq:Liggett} holds. Recall that given parameters $p=1$, $q$ and $\alpha, \beta, \gamma, \delta$, we may define densities $\rho_a, \rho_b$ such that 
\begin{equation} \label{eq:ratestodensity}
    \frac{\alpha}{\rho_a}-\frac{\gamma}{1-\rho_a}=1-q, \quad \quad \frac{\beta}{1-\rho_b}-\frac{\delta}{\rho_b}=1-q. 
\end{equation}
Inspired by the MPA representations found in the TASEP case \cite{derrida1993exact}, let us look for matrices $D$ and $E$ with a bidiagonal structure:
\begin{align}
    D&= \frac{1}{1-q} \sum_{n\geq 1} \hd_n \ket n \bra n + \hdo_n \ket n \bra{n+1}, \\
    E&=\frac{1}{1-q} \sum_{n\geq 1} \he_n \ket n \bra n + \heo_n \ket{n+1} \bra{n}. 
\end{align}
The relation $DE-qED =D+E$ holds if and only if the coefficients satisfy, for all $n\geq 1$, 
\begin{align}
    (1-q)(\hd_n+\he_n)&= \hd_n \he_n + \hdo_n \heo_n -q \hd_n \he_n -q \heo_{n-1}\hdo_{n-1},\label{eq:firstrecurrence}\\ 
    (1-q) \hdo_n &= \hdo_n \he_{n+1} -q \he_n \hdo_n,\\ 
    (1-q) \heo_n &= \hd_{n+1}\heo_n -q\heo_n \hd_n,
\end{align}
with the convention that $\heo_{0}\hdo_{0}=0$. The last two equations simplify to 
\begin{align*}
    \he_{n+1}&=q \he_n+ 1-q \\ 
    \hd_{n+1}&=q \hd_n+ 1-q.
\end{align*} 
These are readily solved as $\he_n=1-e q^{n-1}$ and $\hd_n=1-d q^{n-1}$, where for the moment, $d$ and $e$ are parameters without physical significance. Note that we could here impose that $e=d=0$, to obtain very simple matrices $D$ and $E$. This is essentially the choice made in \cite{sandow1994partially} which is the first reference giving a representation in the general five parameter case. We will see that it is however convenient to keep these two extra degrees of freedom, as in \cite{enaud2004large}. 

Plugging the solution for $\hd_n, \he_n$ in \eqref{eq:firstrecurrence} and letting $\hdo_n \heo_n=\mathsf u_n$, we obtain 
\begin{equation}
\begin{cases}
      \mathsf u_n&=q \mathsf u_{n-1}+ (1-q)(1-ed q^{2n-2}),\\
      \mathsf u_0&=0,
      \end{cases}
\end{equation}
whose unique solution is $\mathsf u_n=(1-q^n)(1-ed q^{n-1})$. At this point, there are several choices possible for $\hdo_n$ and $\heo_n$. Following \cite{enaud2004large}, we may for instance choose that $\hdo_n=1-q^n$ and $\heo_n=1-deq^{n-1}$ for all $n\geq 1$. See \cite{blythe2007nonequilibrium} for a discussion of other choices. 

Now, we need to find the vectors $\bra W$ and $\ket V$. We rewrite \eqref{eq:MPA2} and \eqref{eq:MPA3} as 
\begin{align}
     (\beta D -\delta E) \ket V = \frac{1}{1-q} \left( \frac{\beta}{1-\rho_b} -\frac{\delta}{\rho_b}\right) \ket V,\label{eq:MPAforV}\\
     \bra W (\alpha E-\gamma D) =\frac{1}{1-q} \bra W \left( \frac{\alpha}{\rho_a}-\frac{\gamma}{1-\rho_a}\right).\label{eq:MPAforW}
\end{align}
Let us write $\bra W= \sum_{n\geq 1} w_n \bra n$ and $\ket V = \sum_{n\geq 1} v_n \ket n$. The relation \eqref{eq:MPAforW} implies, for all $n\geq 2$, 
\begin{equation} 
\label{eq:wn}
\alpha w_n \he_n + \alpha w_{n+1} \heo_n - \gamma w_n \hd_n - \gamma w_{n-1} \hdo_{n-1} = \left( \frac{\alpha}{\rho_a}-\frac{\gamma}{1-\rho_a}\right) w_n.  
\end{equation} 
When $n=1$, we obtain the same equation, under the convention that $\hdo_0=0$. 
Given our solutions for $\he_n, \heo_n, \hd_n$ and $\hdo_n$, it yields, for all $n \geq 1$,
\begin{equation}
     (1-e q^{n-1}) w_n+ (1-ed q^{n-1}) w_{n+1} - \frac{\gamma}{\alpha} (1-d q^{n-1}) w_n -\frac{\gamma}{\alpha} (1-q^{n-1})w_{n-1} = \left( \frac{1}{\rho_a} -\frac{\gamma}{\alpha(1-\rho_a)}\right) w_n. 
\end{equation}
This is a three term recursion which, in principle, could be solved using the theory of $q$-deformed orthogonal polynomials -- see \cite{sasamoto1999one, uchiyama2004asymmetric}. However, for well-chosen values of $e$ and $d$, there exists a simple solution. 
Indeed, if we impose 
\begin{equation}
    \label{eq:defe}
    e= \frac{\gamma}{\alpha} \frac{\rho_a}{1-\rho_a}
\end{equation}
one may check that  
\begin{equation} \label{eq:solwn}
	w_n=\left( \frac{1-\rho_a}{\rho_a}\right)^n
\end{equation}
 is a solution of the recursion. Similarly for $\ket V$, the relation \eqref{eq:MPAforV} implies, for $n\geq 2$, 
\begin{equation}
    \beta \hd_n v_n + \beta \hdo_n v_{n+1}-\delta \he_n v_n -\delta \heo_{n-1}v_{n-1}  = \left( \frac{\beta}{1-\rho_b} -\frac{\delta}{\rho_b}\right) v_n.
    \label{eq:vn}
\end{equation}
For $n=1$, we obtain the same relation,  under the convention that $\heo_0=0$. 
Given our solutions for $\hd_n, \hdo_n, \heo_n$, it yields, for $n\geq 2$, 
\begin{equation}
     (1-d q^{n-1}) v_n+ (1-q^n) v_{n+1} - \frac{\delta}{\beta} (1-e q^{n-1}) v_n -\frac{\delta}{\beta} (1-edq^{n-2})v_{n-1} = \left( \frac{1}{1-\rho_b} -\frac{\delta}{\beta\rho_b}\right) v_n, 
\end{equation}
together with a condition for $n=1$. 
If we impose 
\begin{equation} 
\label{eq:defd}
d = \frac{\delta(1-\rho_b)}{\beta\rho_b},
\end{equation}
then one may check that 
\begin{equation} \label{eq:solvn}
	v_n = \left(\frac{\rho_b}{1-\rho_b} \right)^n \frac{(ed;q)_{n-1}}{(q;q)_{n-1}}
\end{equation}
 solves the recursion, where the $q$-Pochhammer symbol $(a;q)_k$ is defined by  $(a;q)_k=(1-a)(1-a)\dots (1-aq^{k-1})$. 
Finally, in view of the relations \eqref{eq:ratestodensity}, \eqref{eq:defe} and \eqref{eq:defd}, imposing $e=d=q$ is equivalent to imposing the condition \eqref{eq:Liggett}, and in that case, we recover the representation given above in \eqref{eq:DEmat} and \eqref{eq:defWV}. 
\begin{remark}
There exists at least one other simple solution for  \eqref{eq:wn} and \eqref{eq:vn}. Letting  $d=\frac{\rho_a}{\rho_a -1}$ and $e=\frac{\rho_b -1}{\rho_b}$, one can see that \eqref{eq:solwn} and \eqref{eq:solvn} are still solutions of the recursion. 
\end{remark}

\bibliographystyle{hyperunsrt}
\bibliography{biblio1.bib} 

\begin{thebibliography}{10}

\bibitem{KPZ}
M.~Kardar, G.~Parisi, and Y.~Zhang.
\newblock Dynamic scaling of growing interfaces.
\newblock {\em Phys. Rev. Lett.}, 56(9):889, 1986.

\bibitem{corwin2016open}
I.~Corwin and H.~Shen.
\newblock Open {ASEP} in the weakly asymmetric regime.
\newblock {\em Comm. Pure Appl. Math.}, 71(10):2065--2128, 2018,
  \href{http://arxiv.org/abs/1610.04931}{{\ttfamily arXiv:1610.04931}}.

\bibitem{borodin2016directed}
A.~Borodin, A.~Bufetov, and I.~Corwin.
\newblock Directed random polymers via nested contour integrals.
\newblock {\em Ann. Phys.}, 368:191--247, 2016,
  \href{http://arxiv.org/abs/1511.07324}{{\ttfamily arXiv:1511.07324}}.

\bibitem{deNardisPLDTT}
J.~De~Nardis, A.~Krajenbrink, P.~Le~Doussal, and T.~Thiery.
\newblock Delta-bose gas on a half-line and the {KPZ} equation: boundary bound
  states and unbinding transitions.
\newblock {\em J. Stat. Mech.: Theor. Exp.}, 2020(4):043207, 2020,
  \href{http://arxiv.org/abs/1911.06133}{{\ttfamily arXiv:1911.06133}}.

\bibitem{corwin2021stationary}
I.~Corwin and A.~Knizel.
\newblock Stationary measure for the open {KPZ} equation.
\newblock {\em arXiv preprint}, 2021,
  \href{http://arxiv.org/abs/2103.12253}{{\ttfamily arXiv:2103.12253}}.

\bibitem{bryc2022markov}
W.~Bryc, A.~Kuznetsov, Y.~Wang, and J.~Weso{\l}owski.
\newblock Markov processes related to the stationary measure for the open {KPZ}
  equation.
\newblock {\em Probab. Theor. Rel. Fields}, pages 1--37, 2022,
  \href{http://arxiv.org/abs/2105.03946}{{\ttfamily arXiv:2105.03946}}.

\bibitem{barraquand2022steady}
G.~Barraquand and P.~Le~Doussal.
\newblock Steady state of the {KPZ} equation on an interval and {Liouville}
  quantum mechanics.
\newblock {\em Europhysics Letters}, 137(6):61003, 2022,
  \href{http://arxiv.org/abs/2105.15178}{{\ttfamily arXiv:2105.15178}}.

\bibitem{goulian1991correlation}
M~Goulian and Miao Li.
\newblock Correlation functions in {L}iouville theory.
\newblock {\em Phys. Rev. Lett.}, 66(16):2051, 1991.

\bibitem{TexierComtetSUSY}
A.~Comtet and C.~Texier.
\newblock One-dimensional disordered supersymmetric quantum mechanics: a brief
  survey.
\newblock {\em Supersymmetry and Integrable Models}, pages 313--328, 1998,
  \href{http://arxiv.org/abs/cond-mat/9707313}{{\ttfamily
  arXiv:cond-mat/9707313}}.

\bibitem{bryc2021markov}
W.~Bryc and A.~Kuznetsov.
\newblock Markov limits of steady states of the {KPZ} equation on an interval.
\newblock {\em arXiv preprint}, 2021,
  \href{http://arxiv.org/abs/2109.04462}{{\ttfamily arXiv:2109.04462}}.

\bibitem{derrida1993exact}
B.~Derrida, M.~R. Evans, V.~Hakim, and V.~Pasquier.
\newblock Exact solution of a {1D} asymmetric exclusion model using a matrix
  formulation.
\newblock {\em J. Phys. A: Math. Gen.}, 26(7):1493, 1993.

\bibitem{bertini1997stochastic}
L.~Bertini and G.~Giacomin.
\newblock Stochastic burgers and {KPZ} equations from particle systems.
\newblock {\em Comm. Math. Phys.}, 183(3):571--607, 1997.

\bibitem{uchiyama2004asymmetric}
M.~Uchiyama, T.~Sasamoto, and M.~Wadati.
\newblock Asymmetric simple exclusion process with open boundaries and
  {Askey--Wilson} polynomials.
\newblock {\em J. Phys. A}, 37(18):4985, 2004,
  \href{http://arxiv.org/abs/cond-mat/0312457}{{\ttfamily
  arXiv:cond-mat/0312457}}.

\bibitem{bryc2010askey}
W{\l}odek Bryc and Jacek Weso{\l}owski.
\newblock {Askey-Wilson} polynomials, quadratic harnesses and martingales.
\newblock {\em Ann. Probab.}, pages 1221--1262, 2010,
  \href{http://arxiv.org/abs/0812.0657}{{\ttfamily arXiv:0812.0657}}.

\bibitem{bryc2017asymmetric}
W.~Bryc and J.~Weso{\l}owski.
\newblock Asymmetric simple exclusion process with open boundaries and
  quadratic harnesses.
\newblock {\em J. Stat. Phys.}, 167(2):383--415, 2017,
  \href{http://arxiv.org/abs/1511.01163}{{\ttfamily arXiv:1511.01163}}.

\bibitem{enaud2004large}
C~Enaud and B~Derrida.
\newblock Large deviation functional of the weakly asymmetric exclusion
  process.
\newblock {\em Journal of statistical physics}, 114(3):537--562, 2004,
  \href{http://arxiv.org/abs/cond-mat/0307023}{{\ttfamily
  arXiv:cond-mat/0307023}}.

\bibitem{derrida2004asymmetric}
B.~Derrida, C.~Enaud, and J.~L. Lebowitz.
\newblock The asymmetric exclusion process and brownian excursions.
\newblock {\em J. Stat. Phys.}, 115(1):365--382, 2004,
  \href{http://arxiv.org/abs/cond-mat/0306078}{{\ttfamily
  arXiv:cond-mat/0306078}}.

\bibitem{derrida2005fluctuations}
B.~Derrida, C.~Enaud, C.~Landim, and S.~Olla.
\newblock Fluctuations in the weakly asymmetric exclusion process with open
  boundary conditions.
\newblock {\em J. Stat. Phys.}, 118(5):795--811, 2005,
  \href{http://arxiv.org/abs/cond-mat/0511275}{{\ttfamily
  arXiv:cond-mat/0511275}}.

\bibitem{brak2006combinatorial}
R.~Brak, S.~Corteel, J.~Essam, R.~Parviainen, and A.~Rechnitzer.
\newblock A combinatorial derivation of the pasep stationary state.
\newblock {\em Electr. J. Comb.}, 13(1):R108, 2006.

\bibitem{blythe2009continued}
R.~A. Blythe, W.~Janke, D.~A. Johnston, and R.~Kenna.
\newblock Continued fractions and the partially asymmetric exclusion process.
\newblock {\em J. Phys. A: Math. Theor.}, 42(32):325002, 2009.

\bibitem{hariya2004limiting}
Y.~Hariya and M.~Yor.
\newblock Limiting distributions associated with moments of exponential
  {B}rownian functionals.
\newblock {\em Stud. Sci. Math. Hung.}, 41(2):193--242, 2004.

\bibitem{barraquand2022stationary}
G.~Barraquand and I.~Corwin.
\newblock Stationary measures for the log-gamma polymer and {KPZ} equation in
  half-space.
\newblock {\em arXiv preprint}, 2022,
  \href{http://arxiv.org/abs/2203.11037}{{\ttfamily arXiv:2203.11037}}.

\bibitem{liggett1975ergodic}
T.~M. Liggett.
\newblock Ergodic theorems for the asymmetric simple exclusion process.
\newblock {\em Trans. Amer. Math. Soc.}, 213:237--261, 1975.

\bibitem{sandow1994partially}
S.~Sandow.
\newblock Partially asymmetric exclusion process with open boundaries.
\newblock {\em Phys. Rev. E}, 50(4):2660, 1994.

\bibitem{essler1996representations}
F.~Essler and V.~Rittenberg.
\newblock Representations of the quadratic algebra and partially asymmetric
  diffusion with open boundaries.
\newblock {\em J. Phys. A: Math. Gen.}, 29(13):3375, 1996.

\bibitem{sasamoto1999one}
T.~Sasamoto.
\newblock One-dimensional partially asymmetric simple exclusion process with
  open boundaries: orthogonal polynomials approach.
\newblock {\em J. Phys. A: Math. Gen.}, 32(41):7109, 1999.

\bibitem{blythe2000exact}
R.~A. Blythe, M.~R. Evans, F.~Colaiori, and F.~Essler.
\newblock Exact solution of a partially asymmetric exclusion model using a
  deformed oscillator algebra.
\newblock {\em J. Phys. A: Math. Gen.}, 33(12):2313, 2000,
  \href{http://arxiv.org/abs/cond-mat/9910242}{{\ttfamily
  arXiv:cond-mat/9910242}}.

\bibitem{blythe2007nonequilibrium}
R.~A. Blythe and M.~R. Evans.
\newblock Nonequilibrium steady states of matrix-product form: a solver's
  guide.
\newblock {\em J. Phys. A: Math. Theor.}, 40(46):R333, 2007,
  \href{http://arxiv.org/abs/0706.1678}{{\ttfamily arXiv:0706.1678}}.

\end{thebibliography}

\end{document}